\documentclass[useAMS,usegraphicx,usenatbib]{mn2e}                      

\usepackage{times}

\newcommand{\Msun}{\ensuremath{\,{\rm M}_\odot}}                        
\newcommand{\Rsun}{\ensuremath{\,{\rm R}_\odot}}                        
\newcommand{\Lsun}{\ensuremath{\,{\rm L}_\odot}}                        
\newcommand{\Teff}{\ensuremath{T_{\rm eff}}}                            
\newcommand{\logg}{\ensuremath{\log g}}                                 
\newcommand{\kms}{\,km\,s$^{-1}$}                                       
\newcommand{\micro}{\ensuremath{v_{\rm turb}}}                          
\newcommand{\Vsync}{\ensuremath{V_{\rm synch}}}                         
\newcommand{\vsini}{\ensuremath{v \sin i}}                              
\newcommand{\Vsini}{\ensuremath{v \sin i}}                              
\newcommand{\Veq}{\ensuremath{V_{\rm eq}}}                              
\newcommand{\Apx}{\,\AA\,px$^{-1}$}                                     
\newcommand{\MoH}{\ensuremath{\left[\frac{\rm M}{\rm H}\right]}}        
\newcommand{\chir}{\ensuremath{\chi_\nu^{\,2}}}                         
\newcommand{\Mbol}{\ensuremath{M_{\rm bol}}}                            
\newcommand{\mc}[1]{\multicolumn{2}{c}{#1}}

\newcommand{\spd}{SPD}
\newcommand{\ion}[2]{{#1}\,{\sc {\small{#2}}}}                          

\newcommand{\reff}[1]{{#1}}                                             

\title[The eclipsing binary YZ\,Cassiopeiae]
      {Absolute dimensions of detached eclipsing binaries. III. The metallic-lined system
YZ\,Cassiopeiae}

\author[Pavlovski et al.]
       {K.\ Pavlovski,$^{1}$ J.\ Southworth,$^{2}$ V.\ Kolbas$^{1}$ and B.\ Smalley$^{2}$ \\
        $^{1}$\,Department of Physics, University of Zagreb, Bijeni\v{c}ka cesta 32, 10000 Zagreb, Croatia \\
        $^{2}$\,Astrophysics Group, Keele University, Newcastle-under-Lyme, ST5 5BG, UK}

\begin{document} \maketitle 

\begin{abstract}
The bright binary system YZ\,Cassiopeiae is a remarkable laboratory for studying the Am phenomenon.
It consists of a metallic-lined A2 star and an F2 dwarf on a circular orbit, which undergo total and
annular eclipses. We present an analysis of 15 published light curves and 42 new high-quality \'echelle spectra,
resulting in measurements of the masses, radii, effective temperatures and photospheric chemical abundances
of the two stars. The masses and radii are measured to 0.5\% precision: $M_{\rm A} = 2.263 \pm 0.012$\Msun,
$M_{\rm B} = 1.325 \pm 0.007$\Msun, $R_{\rm A} = 2.525 \pm 0.011$\Rsun\ and $R_{\rm B} = 1.331 \pm 0.006$\Rsun.
We determine the abundance of 20 elements for the primary star, of which all except scandium are super-solar
by up to 1\,dex. The temperature of this star ($9520 \pm 120$\,K) makes it one of the hottest Am stars.
We also measure the abundances of 25 elements for its companion ($\Teff = 6880 \pm 240$\,K), finding all
to be solar or slighly above solar. The photospheric abundances of the secondary star should be
representative of the bulk composition of both stars. Theoretical stellar evolutionary models are unable
to match these properties: the masses, radii and temperatures imply a half-solar chemical composition
($Z =0.009 \pm 0.003$) and an age of 490--550\,Myr. YZ\,Cas therefore presents a challenge to stellar
evolutionary theory.
\end{abstract}

\begin{keywords}
stars: fundamental parameters --- stars: binaries: eclipsing --- stars: binaries: spectroscopic --- stars:
abundances --- stars: chemically peculiar --- stars: individual: YZ Cas
\end{keywords}


\section{Introduction}

Eclipsing binary star systems are of fundamental importance to astrophysics as they are the primary source
of direct measurements of the physical properties of stars \citep{Andersen91aarv,Torres++10aarv}. The masses
and radii of double-lined systems can be measured to precisions of better than 1\% using only geometric
arguments \citep[e.g.][]{Clausen+08aa}. Obtaining effective temperature (\Teff) measurements converts them
into excellent distance indicators \citep{Ribas+05apj,Bonanos+06apj}. Detached eclipsing binaries (dEBs)
hold an additional advantage as their properties can be used to investigate the predictive abilities of
evolutionary models for single stars.

Metallic-lined A stars (Am stars) are defined to be those which show unusually weak Ca and/or Sc lines
and overly strong metallic lines for their spectral type as derived from Balmer line profiles
\citep{Conti70pasp,GrayCorbally09book}. They were first described as a class of chemically peculiar
stars by \citet{TitusMorgan40apj}. Their abundance anomalies are explained as the product of chemical
stratification caused by radiative levitation and gravitational settling \citep{Michaud70apj,Turcotte+00aa,
Talon++06apj}. Pulsations have recently been found in many Am stars \citep{Smalley+11aa2,Balona+11mn}.

These physical phenomena \reff{can produce observable effects in} quiet radiative atmospheres. The Am phenomenon occurs
only in stars with rotational velocities slower than about 100\kms\ \citep{AbtLevy85apjs,Budaj96aa,Budaj97aa}.
Am stars are preferentially found in binary systems, where tidal effects cause a slower rotation than for
single stars \citep{Abt61apjs,Abt65apjs,CarquillatPrieur07mn}. This means that they are well-represented
in the known population of dEBs \footnote{A catalogue of well-studied detached eclipsing binaries is
maintained at {\tt http://www.astro.keele.ac.uk/$\sim$jkt/debdata/debs.html}}, for example V364\,Lac
\citep{Torres+99aj}, V459\,Cas \citep{Lacy++04aj}, WW\,Cam \citep{Lacy+02aj} and RR\,Lyn
\citep{TomkinFekel06aj}. Whilst the Am phenomenon is a `surface disease', a very high bulk metal
abundance was found for the metallic-lined dEB WW\,Aur by \citet{Me+05mn}.

In this work we present a detailed analysis of the dEB YZ\,Cas, a system which consists of a metallic-lined
A2 star and an F2 dwarf of significantly lower mass, radius and \Teff. This system provides an opportunity
to study the Am phenomenon in a star for which the {\it internal} chemical composition can be inferred --
from its lower-mass companion. YZ\,Cas is particularly well suited to such an analysis. It is totally eclipsing
and extensive photometric \reff{data} is available in the literature, so the radii of the stars can be determined to
unusually high precision. It is also bright and contains slowly rotating stars, making spectroscopic
analysis especially productive. Below we recap the long observational history of this object, model
the available light curves, analyse new \'echelle spectroscopy of the system, deduce the physical
properties of the two stars, and finally confront theoretical models with our results.

\subsection{YZ Cassiopeiae}                                         \label{sec:intro:yzcas}

\begin{table} \begin{center}
\caption{\label{table:yzcasdata} Identifications, location,
and combined photometric indices for YZ\,Cassiopeiae.}
\begin{tabular}{lr@{}lr} \hline
                           &   & YZ Cassiopeiae     & Reference \\ \hline
Flamsteed designation      &   & 21 Cas             & 1   \\
Hipparcos number           &   & HIP 3572           & 2   \\
Henry Draper number        &   & HD 4161            & 3   \\
Bright Star Catalogue      &   & HR 192             & 4   \\
Bonner Durchmusterung      &   & BD +74 27          & 5   \\[2pt]
$\alpha_{2000}$            &   & 00 45 39.078       & 2   \\
$\delta_{2000}$            & + & 74 59 17.06        & 2   \\
Hipparcos parallax ($m$as) &   & 11.24 $\pm$ 0.55   & 2   \\
Spectral type              &   & A2m + F2V          & 6   \\[2pt]
$B_T$                      &   & 5.739 $\pm$ 0.003  & 2   \\
$V_T$                      &   & 5.660 $\pm$ 0.003  & 2   \\
$J_{\rm 2MASS}$            &   & 5.585 $\pm$ 0.019  & 7   \\
$H_{\rm 2MASS}$            &   & 5.644 $\pm$ 0.038  & 7   \\
$K_{\rm 2MASS}$            &   & 5.602 $\pm$ 0.021  & 7   \\
$b-y$                      &   & 0.036 $\pm$ 0.005  & 8   \\
$v-b$                      &   & 0.211 $\pm$ 0.007  & 8   \\
$u-b$                      &   & 1.455 $\pm$ 0.008  & 8   \\
\hline \end{tabular} \end{center}
{\bf References:} (1) \citet{flamsteed}; (2) \citet{Perryman+97aa};
(3) \citet{CannonPickering18anhar2}; (4) \citet{HoffleitJaschek91};
(5) \citet{Argelander03book}; (6) This work; (7) 2MASS \citep{Cutri+03book};
(8) \citet{HilditchHill75mmras}, given as the mean and standard deviation of the
six measurements for each Str\"omgren colour index (all taken outside eclipse).
\end{table}

YZ\,Cas shows total and annular eclipses recurring on an orbital period of 4.47\,d. It has been the subject
of many analyses over nearly a century, which we summarise below. Throughout this work we refer to the
primary component as star\,A and the secondary component as star\,B. Star\,A is substantially hotter,
larger and more massive than star\,B, and is the component in inferior conjunction at the midpoint of
primary eclipse. The spectral type of the YZ\,Cas system is given as A2 in the {\it Henry Draper Catalogue}
\citep{CannonPickering18anhar2}; this measurement pertains to star\,A, which dominates the light of the system.

The eclipsing nature of YZ\,Cas was announced by \citet{Stebbins24pa}, who is credited with the discovery,
and further elaborated by \citet{Huffer25pa}. It was chosen at the Washburn Observatory as a photometric
 comparison star for 23\,Cas (a spectroscopic binary which has not been observed to be variable).
The resulting light curve was presented by \citet{Huffer28pwo}.

\citet{Plaskett26pdao} determined the first spectroscopic orbit of the system, based on radial velocity
(RV) measurements of star\,A. Under the assumption that both stars had a mass of 2\Msun\ he measured
their radii and orbital separation. Plaskett detected the Rossiter-McLaughlin effect
\citep{Rossiter24apj,McLaughlin24apj} during primary minimum, with an amplitude slightly greater
than 3\kms. The shape of the Rossiter-McLaughlin anomaly is consistent with alignment between the
orbital and stellar rotational axes, albeit with low significance.

\citet{Kron39pasp} used the annular nature of secondary eclipse to measure the colour index of star\,B,
\reff{based on photoelectric
photometry from \citet{Huffer31pwo},} and transformed this into an approximate spectral type of F4.

\citet{Kron39licob2} presented and tabulated a light curve of the system with full coverage of the
eclipses, obtained using a photoelectric photometer with an effective wavelength of 4500\,\AA\
\citep{Kron39licob1}. He used this to measure the fractional radii and limb darkening (LD) coefficients
of the two stars \citep[see also][]{Kron38pasp}. \citet{Kron42apj} used the same instrument but with
a red filter and red-sensitive light detector to obtain a second light curve in a passband covering
5200--8200\,\AA\ (half maximum response). He found fractional radii in good agreement with those from
his blue light curve, and determined a colour of star\,B corresponding to a spectral type of F5.
\citet{Serkowski61aj} reconsidered the determination of the LD coefficients from the Kron light curves.

\citet{Mcnamara51phd} presented two new light curves in a blue and a UV passband, obtained using
similar equipment and methods to \citet{Kron39licob2}. The main aim was to investigate
the LD at blue wavelengths.
\citet{Grygar++77baicz} presented a detailed discussion of LD from the two Kron and the two McNamara
light curves, and revised measurements of the physical properties of the stars.

\citet{Koch++65apj} measured the spectral type of star\,A to be A2-A3 and its equatorial rotational
velocity to be $\Veq = 34 \pm 2$\kms. \citet{PerryStone66pasp} obtained a single-lined
spectroscopic orbit from 92 RVs. Their \reff{measured} orbital eccentricity, $e = 0.004 \pm 0.004$
(most likely a probable rather than standard error), suggests that the orbit is circular \citep{LucySweeney71aj}.

\citet{Lacy81apj} \reff{studied} YZ\,Cas using the Kron light curves and new spectra from
which RVs were measured for star\,B for the first time. \citeauthor{Lacy81apj} adopted photometric parameters
which were the weighted average of eight independent studies of the same light curves
\citep{Kron39licob2,Kron42apj}, a procedure which may lack statistical validity. Lacy also obtained
individual \reff{Str\"omgren} photometric indices of the two component stars from published
photometry. Lacy's RVs yielded measurements of the masses of the stars to a precision of 0.5\%,
allied with values of $\Veq$ of $34 \pm 1$\kms\ for star\,A and $16 \pm 2$\kms\ for star\,B.

Shortly after publication of the study of \citet{Lacy81apj}, \citet{Delandtsheer83aas} presented a large
number of photoelectric observations of YZ\,Cas. These were obtained in four colours in the Utrecht photometric
system \citep{Provoost80aas,Heintze89ssrv}, and were analysed using the code of \citet{WilsonDevinney71apj}
 in order to obtain the physical properties of the stars. \citet{DelandtsheerMulder83aa} obtained two
high-resolution UV spectra from the {\it International Ultraviolet Explorer} satellite and confirmed
the metallic nature of star\,A. The abundance of star\,B was not measured, so the
internal (rather than photospheric) abundance of star\,A remained unknown. \citet{Delandtsheerdegreve84aa}
found a subsolar metal abundance ($Z=0.015$ where $Z$ is the mass fraction of metals) in a comparison
with theoretical stellar evolutionary models, in \reff{disagreement} with Lacy's finding of a supersolar bulk
metal abundance of $Z = 0.027 \pm 0.003$.

Finally, \citet{Papousek89pbrno} obtained 16\,000 photoelectric observations in six passbands,
from the 60\,cm reflector of the University Observatory, Brno. He used these to study the physical
properties and LD of the system.


\section{Observations and data reduction}

\subsection{Spectroscopy}

We obtained 42 high-resolution spectra of YZ\,Cas in October 2007, using the Nordic Optical Telescope (NOT)
at La Palma, Spain, equipped with the FIbre-fed Echelle Spectrograph (FIES). This spectrograph is housed
in a \reff{separate climate-controlled building} and has a high thermal and mechanical stability.
The wavelength scale was established from thorium-argon exposures taken regularly throughout the observing
nights. We used fibre 4 in fibre bundle B, giving complete spectral coverage in the interval 3640--7360\,\AA\
at a reciprocal dispersion ranging from 0.023\Apx\ in the blue to 0.045\Apx\ in the red. The resolution of
the instrument is roughly 3.5\,px, giving a resolving power of 48\,000. An exposure time of 300\,s was used
for all spectra, resulting in continuum signal to noise ratios in the region of 300 in the $B$ and $V$ bands.

The basic steps for data reduction (bias subtraction, flat-fielding, correction for scattered light, extraction
of orders, and wavelength calibration) were performed with IRAF\footnote{IRAF is distributed by the National
Optical Astronomy Observatory, which is operated by the Association of Universities for Research in Astronomy
(AURA) under cooperative agreement with the National Science Foundation.}. Removal of the instrumental blaze
function for YZ\,Cas is not trivial, because the broad Balmer lines from star\,A extend over entire \'echelle
orders. In such cases we interpolated blaze functions from adjacent orders which are well-defined, using
a semi-manual approach and {\sc java} routines written by VK.

\subsection{Photometry}

\begin{table} \begin{center}
\caption{\label{tab:passbands} The wavelengths of maximum transmission
($\lambda_{\rm cen}$) and the full widths at half maximum response (FWHM) of the
passbands of the light curves used in this work. When possible the FWHM is given
as the actual wavelength interval rather than just the width of the passband.}
\begin{tabular}{lccl} \hline
Passband & $\lambda_{\rm cen}$ (\AA) & FWHM (\AA) & Reference \\
\hline
Kron blue       & 4500 &   (wide)   & \citet{Kron39licob1}    \\
Kron red        & 6000 & 5250--8220 & \citet{Kron42apj}       \\
McNamara UV     & 3575 &    850     & \citet{Mcnamara51phd}   \\
McNamara blue   & 4525 & 3500--5700 & \citet{Mcnamara51phd}   \\
Utrecht 472     & 4730 &    105     & \citet{Provoost80aas}   \\
Utrecht 672     & 6719 &    100     & \citet{Provoost80aas}   \\
Utrecht 782     & 7812 &    115     & \citet{Provoost80aas}   \\
Utrecht 871     & 8798 &    130     & \citet{Provoost80aas}   \\
U1              & 3500 &    260     & \citet{Papousek89pbrno} \\
U2              & 3810 &    155     & \citet{Papousek89pbrno} \\
B               & 4075 &    190     & \citet{Papousek89pbrno} \\
G               & 4903 &    100     & \citet{Papousek89pbrno} \\
V               & 5405 &    145     & \citet{Papousek89pbrno} \\
O               & 5822 &    100     & \citet{Papousek89pbrno} \\
{\it Hipparcos} & 4800 & 4280--6550 & \citet{Bessell00pasp}   \\
\hline \end{tabular} \end{center} \end{table}

We present no new photometry in this work, due to the large number and good quality of existing datasets.
Published observations were taken from a range of sources, discussed in Sect.\,\ref{sec:intro:yzcas},
resulting in a total of 15 separate light curves. All of these were observed in non-standard passbands,
so we give a summary of them in Table\,\ref{tab:passbands}. We did not use the data from \citet{Huffer28pwo}
as they are sparse and relatively scattered.

\citet{Kron39licob2} and \citet{Kron42apj} presented photoelectric photometry of YZ\,Cas in one blue and
one red wide passband, obtained using a 1\,m reflector at the Lick Observatory. The two Kron light curves
were obtained from the papers by optical character recognition using the
{\sc tesseract}\footnote{\tt http://code.google.com/p/tesseract-ocr/} software. Due to changes in the
optical path of the instrument between the variable star and its comparison star (23\,Cas), many of the
nights of data were shifted in magnitude to obtain a good internal agreement. Becase of this, the
observations taken well outside eclipse carry essentially no information. We have therefore rejected
data taken more than 0.05 phase units from the midpoint of an eclipse. A few observations were enclosed
in square brackets to indicate that they are less reliable, and we also rejected these. Each Kron datapoint
is the sum of six individual deflections on the chart recorder, meaning they have a relatively high precision.

\citet{Mcnamara51phd} observed YZ\,Cas in one UV and one blue passband, with central wavelengths of 3575\,\AA\
and 4525\,\AA\ for an A0 star. The UV passband was selected using a Corning O-5840 glass filter, and the blue
passband with Corning C-5562 and O-3389 glass. The passband widths were not specified by \citet{Mcnamara51phd},
so have been obtained from diagrams in \citet{Dobrowolski++77ao} and are given in Table\,\ref{tab:passbands}.
The blue passband has more than 50\% transmission from $4300 \pm 50$\,\AA\ to $5700 \pm 150$\,\AA, but a long
red tail of roughly 30\% transmission extends beyond the edge of Fig.\,20 in \citet{Dobrowolski++77ao} at
7500\,\AA. The photometric data were obtained from \citet{Mcnamara51phd} as a function of orbital phase.

\citet{Delandtsheer83aas} obtained light curves of YZ\,Cas in four different narrow passbands in the
Utrecht photometric system, centred on wavelengths 474, 672, 782 and 871 nm. These were aimed at wavelength
intervals containing no telluric lines and the fewest stellar spectral lines possible, with the motivation
of improving the understanding of continuum LD through study of dEBs. Only one set of Utrecht filters is
known to exist, and these were affixed to a 40\,cm reflector then operated at Ausserbin in the Swiss Alps.
The four Utrecht light curves had to be obtained from the paper by copying out by hand. They have a much
larger scatter than the Kron data, but the morphology of the light variation of YZ\,Cas means they are
still valuable data.

YZ\,Cas was bright enough to be observed by the {\it Hipparcos} satellite \citep{Perryman+97aa} and also
the {\it Tycho} experiment on board {\it Hipparcos} \citep{Hog+97aa}. All three datasets suffer from
a shortage of points within the total phases of secondary minimum, but in the case of the {\it Hipparcos}
passband the scatter is sufficiently small that the points on the ascending branch of the secondary
minimum provide adequate constraints on the depth of the minimum.

\citet{Papousek89pbrno} presented possibly the most extensive observations of YZ\,Cas. These comprise
about 16\,000 observations in an intermediate-band filter system denoted $U1$, $U2$, $B$, $G$, $V$ and $O$,
in order of increasing wavelength. These are tabulated in his paper as sets of normal points covering phases
between 0 and 1, and were converted into machine-readable format using {\sc tesseract}. As they are specified
as magnitudes versus phase we do not have information on the times of individual observations, only that
the data were obtained in the years 1973--1974.


\section{Orbital period determination}                                                     \label{sec:porb}

\begin{table} \begin{center}
\caption{\label{tab:minima} Literature times of minimum light of YZ\,Cas and the observed
minus calculated ($O-C$) values of the data compared to the ephemeris derived in this work.}
\begin{tabular}{lrrr} \hline
Time of minimum          &  Cycle  & $O-C$ value & Reference \\
(HJD $-$ 2\,400\,000)    &  number & (HJD)       &           \\
\hline
23716.7318 $\pm$ 0.01       & $-$4895.0 & $-$0.00136 &  1  \\     
23966.8931 $\pm$ 0.005      & $-$4839.0 & $-$0.00452 &  2  \\     
23975.8282 $\pm$ 0.003      & $-$4837.0 & $-$0.00386 &  3  \\     
25374.070 $\pm$ 0.005       & $-$4524.0 & $-$0.00266 &  1  \\     
25414.274 $\pm$ 0.01        & $-$4515.0 & $-$0.00366 &  1  \\     
25512.568 $\pm$ 0.01        & $-$4493.0 &    0.01144 &  4  \\     
25776.122 $\pm$ 0.01        & $-$4434.0 & $-$0.00068 &  4  \\     
26562.354 $\pm$ 0.01        & $-$4258.0 &    0.00019 &  4  \\     
26754.443 $\pm$ 0.005       & $-$4215.0 & $-$0.00137 &  5  \\     
26763.386 $\pm$ 0.005       & $-$4213.0 &    0.00718 &  5  \\     
26946.5317 $\pm$ 0.005      & $-$4172.0 & $-$0.00324 &  6  \\     
26955.4548 $\pm$ 0.005      & $-$4170.0 & $-$0.01458 &  6  \\     
26973.357 $\pm$ 0.005       & $-$4166.0 &    0.01873 &  5  \\     
28733.4218 $\pm$ 0.003      & $-$3772.0 & $-$0.00208 &  7  \\     
29175.6774 $\pm$ 0.005      & $-$3673.0 & $-$0.00150 &  8  \\     
41355.5615 $\pm$ 0.0019     &  $-$946.5 &    0.00082 &  9  \\     
41880.45922 $\pm$ 0.00020   &  $-$829.0 & $-$0.00009 & 10  \\     
42148.49266 $\pm$ 0.00010   &  $-$769.0 &    0.00001 & 10  \\     
43256.3644 $\pm$ 0.0003     &  $-$521.0 &    0.00060 &  9  \\     
44632.2678 $\pm$ 0.0005     &  $-$213.0 & $-$0.00049 &  9  \\     
44632.2685 $\pm$ 0.0010     &  $-$213.0 &    0.00021 & 11  \\     
44929.3378 $\pm$ 0.0009     &  $-$146.5 & $-$0.00078 &  9  \\     
45583.7863 $\pm$ 0.0009     &       0.0 & $-$0.00035 & 12  \\     
45583.7867 $\pm$ 0.0004     &       0.0 &    0.00005 & 12  \\     
45621.748 $\pm$ 0.005       &       8.5 & $-$0.01004 & 12  \\     
45621.756 $\pm$ 0.004       &       8.5 & $-$0.00204 & 12  \\     
45990.297 $\pm$ 0.004       &      91.0 & $-$0.00689 & 12  \\     
45990.305 $\pm$ 0.006       &      91.0 &    0.00111 & 12  \\     
48411.53820 $\pm$ 0.00056   &     633.0 & $-$0.00020 & 13  \\     
48411.5384 $\pm$ 0.0032     &     633.0 &    0.00000 & 13  \\     
48411.5372 $\pm$ 0.0030     &     633.0 & $-$0.00120 & 13  \\     
48822.5201 $\pm$ 0.003      &     725.0 & $-$0.00277 & 14  \\     
54509.2969 $\pm$ 0.0003     &    1998.0 & $-$0.00004 & 15  \\     
\hline \end{tabular} \end{center}
{\bf References:} (1) \citet{Kukarkin28pz}; (2) \citet{Plaskett26pdao}; (3) \citet{Huffer31pwo};
(4) \citet{Zverev36sp}; (5) \citet{Hassenstein54popot}; (6) \citet{Skoberla36za};
(7) \citet{Kron39licob2}; (8) \citet{Dombrovskij64trlen}; (9) \citet{LavrovLavrova88trkaz};
(10) \citet{Papousek89pbrno}; (11) \citet{Delandtsheer83aas}; (12) \citet{DiethelmLines86ibvs};
(13) This work, based on the {\it Hipparcos} and {\it Tycho} $B$ and $V$ observations;
(14) \citet{Brelstaff94baa}; (15) P.\ Svoboda (in \citealt{Brat+08oejv}).
\end{table}

The available times of minimum light for YZ\,Cas extend from 1923 to 2008. Most of these are listed by
\citet{Kreiner++01book} and \citet{Papousek89pbrno}, to which we added a few additional values from
a literature search. We rejected several recent visual timings which were either discrepant or too
uncertain to be useful. A straight line was fitted to these timings as a function of HJD, from which
we determined the orbital ephemeris:
\[ {\rm Min\,I} = {\rm HJD}\ 2\,445\,583.78664 (10) + 4.46722236 (11) \times E \]
The individual times of minimum light are given in Table\,\ref{tab:minima} along with their references
and residuals versus the fitted ephemeris. The time system used in this work is UTC, which does not
compensate for leap seconds; the time of primary eclipse is only specified to within 9\,s so this does
not have a significant effect on the results.

A small number of timings refer to secondary rather than primary minimum. We assumed these to represent
phase 0.5, and their residuals support this assumption. The minimum timings are therefore consistent
with a circular orbit.


\section{Light curve analysis}                                                                   \label{sec:lc}

The 15 available light curves of YZ\,Cas have been individually modelled using the {\sc jktebop} code%
\footnote{{\sc jktebop} is written in {\sc fortran77} and the source code is available at
{\tt http://www.astro.keele.ac.uk/jkt/}}
code \citep{Me++04mn,Me13aa}, which is based on the {\sc ebop} program developed by
P.\ B.\ Etzel \citep{PopperEtzel81aj,Etzel81conf,NelsonDavis72apj}. This code represents the stellar
figures as biaxial spheroids\reff{,} with the spherical approximation used for the calculation of eclipse
functions. This approximation is more than adequate for well-detached EBs such as YZ\,Cas. Tests for
third light and orbital eccentricity returned small and insignificant values, so these parameters were
held at zero for the final solutions. The orbital period was fixed at the value found in
Sect.\,\ref{sec:porb}, but the ephemeris zeropoint was included as a fitted parameter. The brightness
outside eclipse was fitted for, as were the sum and ratio of the fractional radii ($r_{\rm A}+r_{\rm B}$
and $k$), the orbital inclination ($i$), and the central surface brightness ratio of the stars ($J$).

YZ\,Cas has a rich history of use for measuring the LD of intermediate-mass stars. For the Kron and
Utrecht light curves we were able to fit for the linear LD coefficient for star\,A ($u_{\rm A}$) but
not for star\,B ($u_{\rm B}$). We therefore fixed $u_{\rm B}$ to theoretical values
\citep{Vanhamme93aj,Claret00aa,Claret04aa2,ClaretHauschildt03aa} appropriate for the photometric
passband used. There is mounting observational evidence that the more complex two-parameter LD laws
can provide a significant improvement \citep{Me++07aa,Me++07mn,Me+09mn2}, \reff{so we tried the quadratic and logarithmic laws.
The use of these laws did not yield better fits, so we adopted} the linear law
for our final results. This is in line with the experience of others when analysing dEBs \citep[e.g.][]{Lacy+10aj}.

Uncertainties in each solution were found using 10\,000 Monte Carlo simulations \citep{Me++04mn2,Me+05mn}.
 $u_{\rm B}$ was fixed at theoretical values when finding the best fits but perturbed by $\pm$0.1 on a flat
distribution during the Monte Carlo simulations, to account for our imperfect knowledge of this quantity.
We also performed a residual permutation analysis \citep{Me08mn} in order to account for possible correlated
noise, and retained those errorbars which were larger than the Monte Carlo ones.

The uncertainties in the resulting photometric parameters are low, due in particular to the primary
eclipse being total and the secondary eclipse annular. Inspection of the Monte Carlo and residual
permutation results reveals two important correlations. One is between $r_{\rm A}$ and $i$, with
a correlation coefficient of $-0.88$, and arises because these parameters govern the eclipse duration.
The other is between $k$ and $u_{\rm A}$, with a correlation coefficient of $-0.93$, and is expected
as the two parameters affect the depth of primary minimum.

For the Kron data we noticed a \reff{systematic mismatch between the data and the best fit} immediately
before and after
the minima. We therefore set the coefficients of the reflection effect to be fitted parameters,
in effect allowing for different normalisations for primary and secondary eclipse. This approach
yielded a significantly better fit, but had a negligible effect on the values of the derived parameters.

\begin{figure*} \includegraphics[width=\textwidth,angle=0]{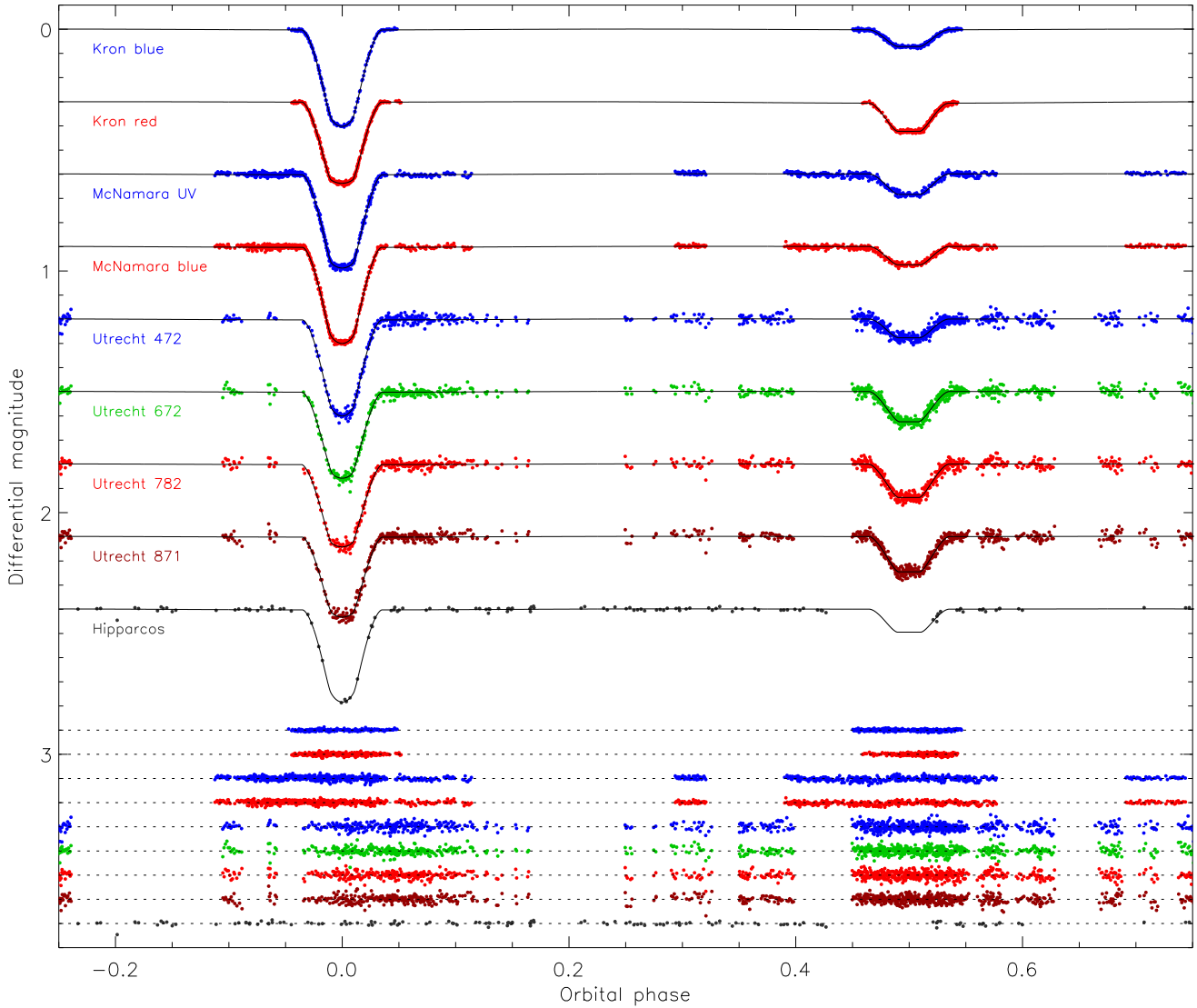}
\caption{\label{fig:plotlc} Light curves of YZ\,Cas from \citet{Kron39licob2},
\citet{Kron42apj}, \citet{Mcnamara51phd}, \citet{Delandtsheer83aas} and {\it Hipparcos}
(coloured points) compared to the {\sc jktebop} best fits (black curves). The residuals
of the fits are offset from zero to appear at the base of the figure.} \end{figure*}

\begin{figure*} \includegraphics[width=\textwidth,angle=0]{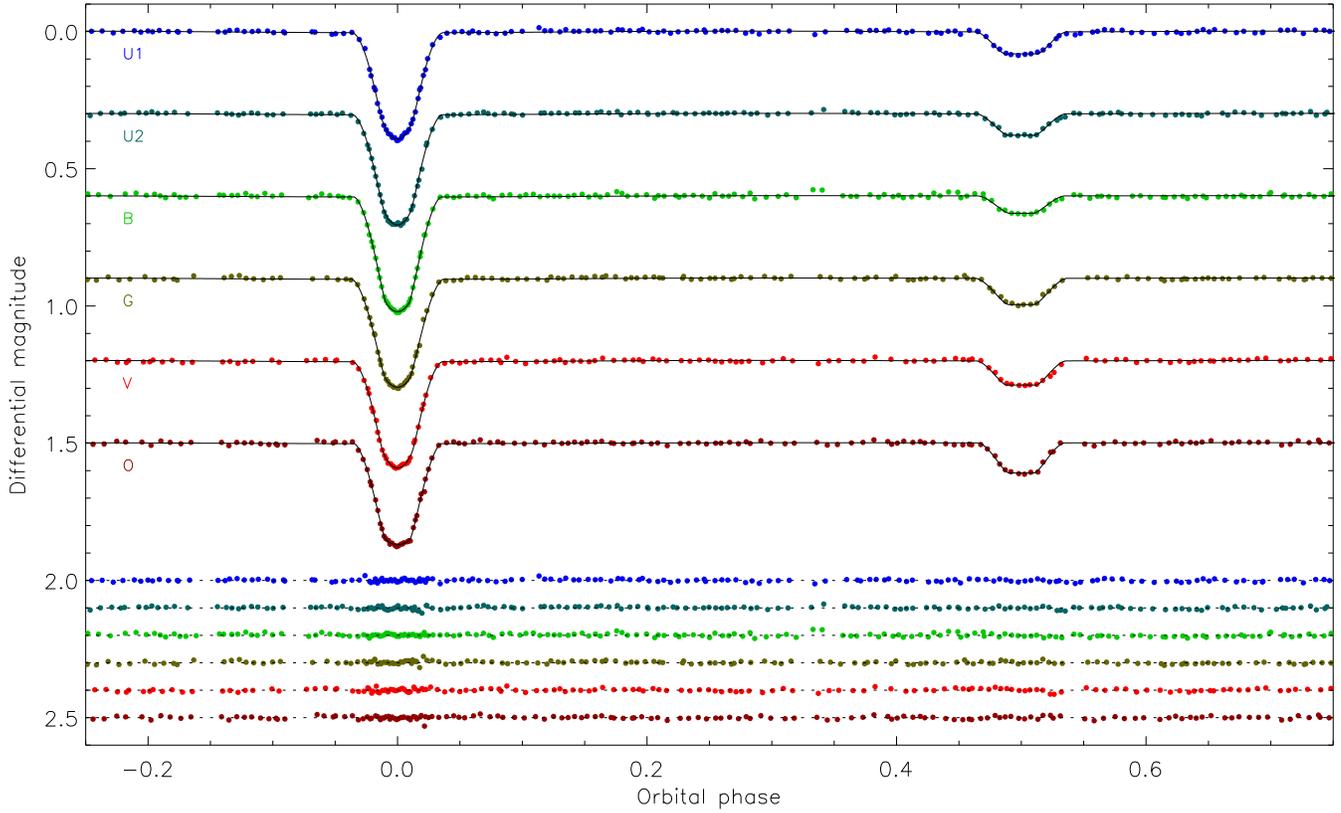}
\caption{\label{fig:plotlcpap} Light curves of YZ\,Cas from \citet{Papousek89pbrno}
(coloured points) compared to the {\sc jktebop} best fits (black curves). The residuals
of the fits are offset from zero to appear at the base of the figure.} \end{figure*}

The fitted light curves are shown in Figs.\ \ref{fig:plotlc} and \ref{fig:plotlcpap}, and the parameters
of the fitted models in Tables \ref{tab:lcfit}, \ref{tab:lcfitpap} and \ref{tab:lcfit3}. The parameters
of the fits unfortunately are in poor agreement with each other, suggesting that the Monte Carlo and
residual-permutation uncertainties are underestimated. It is also the case that the two light curves
which yield the most precise parameters (the Kron datasets) suffer from systematic errors due to optical
path changes in the instrument, so their uncertainties have an additional reason to be underestimated.
The light curves with the greatest discrepancy with respect to the others is $U1$ from \citet{Papousek89pbrno},
which has the bluest passband of all datasets studied in the current work. This light curve was observed
from a low-altitude site in mainland Europe. Our experience of such observatories is that the atmosphere
is almost always unstable on hourly timescales, leading to extinction variations and thus systematic errors.
These are particularly pronounced in the blue, where atmospheric extinction is greater.

\begin{table*} \begin{center} \caption{\label{tab:lcfit} Model parameters of the Kron and
Utrecht light curves of YZ\,Cas. The upper part of the table contains fitted parameters
and the lower part contains derived parameters. $N_{\rm obs}$ is the number of datapoints
in each light curve and the scatter is the rms of the residuals around the best fit.}
\begin{tabular}{l r@{\,$\pm$\,}l r@{\,$\pm$\,}l r@{\,$\pm$\,}l r@{\,$\pm$\,}l r@{\,$\pm$\,}l r@{\,$\pm$\,}l r@{\,$\pm$\,}l} \hline
Parameter         &   \mc{Kron blue}    &   \mc{Kron red}     &  \mc{Utrecht 472}   &  \mc{Utrecht 672}   &  \mc{Utrecht 781}   &  \mc{Utrecht 871}   \\
\hline
$J$               &  0.229 & 0.006      &    0.441 & 0.012    &    0.255 & 0.010    &    0.421 & 0.014    &    0.476 & 0.015    &    0.503 & 0.015    \\
$r_{\rm A}+r_{\rm B}$&0.2217 & 0.0009   &   0.2178 & 0.0008   &   0.2198 & 0.0029   &   0.2245 & 0.0026   &   0.2235 & 0.0025   &   0.2224 & 0.0025   \\
$k$               &   0.5286 & 0.0025   &   0.5166 & 0.0017   &   0.5345 & 0.0080   &   0.5310 & 0.0054   &   0.5285 & 0.0051   &   0.5215 & 0.0050   \\
$u_{\rm A}$       &    0.580 & 0.029    &    0.361 & 0.023    &    0.531 & 0.092    &    0.460 & 0.078    &    0.359 & 0.076    &    0.392 & 0.073    \\
$u_{\rm B}$       & \mc{0.65 perturbed} & \mc{0.50 perturbed} & \mc{0.65 perturbed} & \mc{0.50 perturbed} & \mc{0.40 perturbed} & \mc{0.35 perturbed} \\
$i$ (\degr)       &    88.18 & 0.10     &    88.36 & 0.10     &    88.18 & 0.39     &    87.49 & 0.25     &    87.84 & 0.29     &    87.87 & 0.30     \\
\hline
$r_{\rm A}$       &   0.1450 & 0.0007   &   0.1436 & 0.0006   &   0.1432 & 0.0021   &   0.1467 & 0.0017   &   0.1462 & 0.0017   &   0.1461 & 0.0017   \\
$r_{\rm B}$       &   0.0767 & 0.0003   &   0.0742 & 0.0003   &   0.0766 & 0.0012   &   0.0779 & 0.0011   &   0.0773 & 0.0010   &   0.0762 & 0.0010   \\
Light ratio       &   0.0621 & 0.0006   &   0.1112 & 0.0008   &   0.0692 & 0.0016   &   0.1166 & 0.0017   &   0.1308 & 0.0018   &   0.1387 & 0.0018   \\
$N_{\rm obs}$     &      \mc{ 407}      &      \mc{ 565}      &      \mc{ 827}      &      \mc{ 829}      &      \mc{ 835}      &      \mc{ 833}      \\
Scatter (mmag)    &      \mc{ 3.9}      &      \mc{ 5.0}      &      \mc{14.7}      &      \mc{14.1}      &      \mc{14.3}      &      \mc{14.6}      \\
\hline \end{tabular} \end{center} \end{table*}

\begin{table*} \begin{center} \caption{\label{tab:lcfitpap} Model parameters of the
\citet{Papousek89pbrno} light curves of YZ\,Cas. See Table\,\ref{tab:lcfit} for other details.}
\begin{tabular}{l r@{\,$\pm$\,}l r@{\,$\pm$\,}l r@{\,$\pm$\,}l r@{\,$\pm$\,}l r@{\,$\pm$\,}l r@{\,$\pm$\,}l r@{\,$\pm$\,}l} \hline
Parameter         &      \mc{$U1$}      &      \mc{$U2$}      &      \mc{$B$}       &      \mc{$G$}       &      \mc{$V$}       &      \mc{$O$}       \\
\hline
$J$               &   0.2771 & 0.0103   &   0.2584 & 0.0098   &   0.2020 & 0.0095   &   0.3049 & 0.0119   &   0.2921 & 0.0127   &   0.3777 & 0.0139   \\
$r_{\rm A}+r_{\rm B}$&0.2285 & 0.0030   &   0.2280 & 0.0019   &   0.2222 & 0.0028   &   0.2274 & 0.0022   &   0.2231 & 0.0022   &   0.2192 & 0.0021   \\
$k$               &   0.4910 & 0.0084   &   0.5260 & 0.0084   &   0.5317 & 0.0091   &   0.5249 & 0.0076   &   0.5126 & 0.0073   &   0.5409 & 0.0059   \\
$u_{\rm A}$       &   0.892  & 0.073    &   0.666  & 0.085    &   0.638  & 0.088    &   0.639  & 0.074    &   0.688  & 0.071    &   0.320  & 0.078    \\
$u_{\rm B}$       & \mc{0.70 perturbed} & \mc{0.70 perturbed} & \mc{0.65 perturbed} & \mc{0.65 perturbed} & \mc{0.60 perturbed} & \mc{0.55 perturbed} \\
$i$ (\degr)       &  89.96   & 0.65     &  88.18   & 0.29     &  88.58   & 0.41     &  88.46   & 0.34     &  89.11   & 0.66     &  88.60   & 0.36     \\
\hline
$r_{\rm A}$       &   0.1533 & 0.0026   &   0.1494 &  0.0018  &   0.1451 & 0.0022   &   0.1491 & 0.0018   &   0.1475 & 0.0018   &   0.1423 & 0.0016   \\
$r_{\rm B}$       &   0.0752 & 0.0009   &   0.0786 &  0.0010  &   0.0771 & 0.0009   &   0.0783 & 0.0010   &   0.0756 & 0.0009   &   0.0770 & 0.0008   \\
Light ratio       &   0.0724 & 0.0019   &   0.0703 &  0.0020  &   0.0567 & 0.0030   &   0.0835 & 0.0021   &   0.0796 & 0.0031   &   0.1009 & 0.0023   \\
$N_{\rm obs}$     &      \mc{175}       &      \mc{175}       &      \mc{183}       &      \mc{164}       &      \mc{158}       &      \mc{164}       \\
Scatter (mmag)    &      \mc{4.7}       &      \mc{4.5}       &      \mc{5.6}       &      \mc{4.9}       &      \mc{5.2}       &      \mc{5.2}       \\
\hline \end{tabular} \end{center} \end{table*}

\begin{table*} \begin{center} \caption{\label{tab:lcfit3} Model parameters of the \citet{Mcnamara51phd}
and {\it Hipparcos} light curves of YZ\,Cas. The final parameter values, obtained from all fifteen light
curves, are given in the last column. See Table\,\ref{tab:lcfit} for other details.}
\begin{tabular}{l r@{\,$\pm$\,}l r@{\,$\pm$\,}l r@{\,$\pm$\,}l r@{\,$\pm$\,}l} \hline
Parameter         &   \mc{McNamara UV}  &  \mc{McNamara blue}  & \mc{\it Hipparcos} &\mc{\bf Final values}\\
\hline
$J$               &    0.252 & 0.010    &    0.293 & 0.012     &   0.321 & 0.075    &       \mc{ }        \\
$r_{\rm A}+r_{\rm B}$&0.2179 & 0.0014   &   0.2201 & 0.0017    &  0.2197 & 0.0053   &  0.22084 & 0.00077  \\
$k$               &   0.5346 & 0.0054   &   0.5422 & 0.0050    &   0.521 & 0.032    &   0.5246 & 0.0026   \\
$u_{\rm A}$       &   0.494  & 0.055    &   0.354  & 0.064     &    0.62 & 0.29     &       \mc{ }        \\
$u_{\rm B}$       & \mc{0.65 perturbed} & \mc{0.65 perturbed}  &\mc{0.60 perturbed} &       \mc{ }        \\
$i$ (\degr)       &  88.81   & 0.26     &  88.51   & 0.17      &   88.49 & 0.97     &   88.332 & 0.066    \\
\hline
$r_{\rm A}$       &   0.1420 & 0.0011   &   0.1427 & 0.0013    &  0.1445 & 0.0062   &  0.14456 & 0.00056  \\
$r_{\rm B}$       &   0.0759 & 0.0006   &   0.0774 & 0.0006    &  0.0752 & 0.0023   &  0.07622 & 0.00033  \\
Light ratio       &   0.0674 & 0.0020   &   0.0764 & 0.0021    &   0.088 & 0.019    &       \mc{ }        \\
$N_{\rm obs}$     &      \mc{894}       &      \mc{893}        &     \mc{ 124}      &       \mc{ }        \\
Scatter (mmag)    &      \mc{6.6}       &      \mc{7.4}        &     \mc{ 7.0}      &       \mc{ }        \\
\hline \end{tabular} \end{center} \end{table*}

\begin{figure} \includegraphics[width=\columnwidth,angle=0]{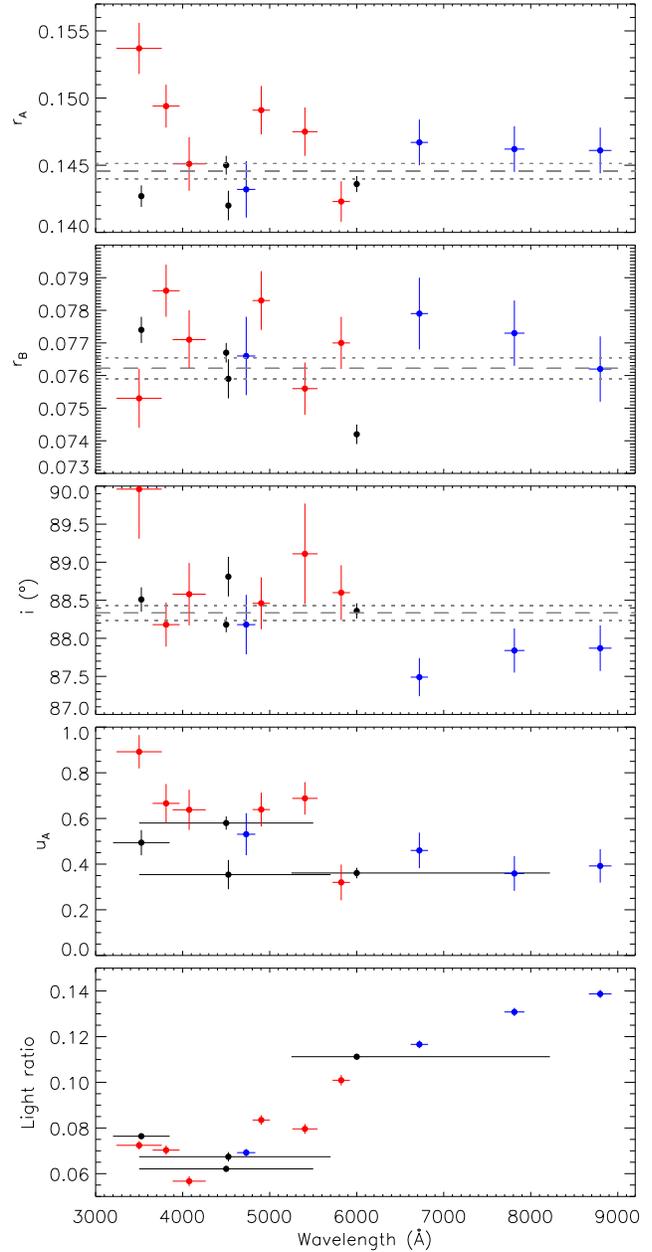}
\caption{\label{fig:lcwave} The photometric parameters $r_{\rm A}$, $r_{\rm B}$,
$i$, $u_{\rm A}$ and light ratio as a function of wavelength. The Utrecht results
are shown using \reff{blue} lines, the \citet{Papousek89pbrno} results using
\reff{red} lines, \reff{and other results using black lines.}
The horizontal lines show the widths of the passbands, but the wide passbands are
indicated only for the wavelength-dependent quantities in the lower two panels. The
dashed horizontal lines show the adopted final values and the dotted horizontal
lines indicate the size of the errorbars on these quantities.} \end{figure}

Fig.\,\ref{fig:lcwave} shows the photometric parameters $r_{\rm A}$, $r_{\rm B}$, $i$, $u_{\rm A}$ and light ratio
as a function of wavelength. YZ\,Cas has sometimes in the past been taken to indicate possible variations of radius
with wavelength. Aside from the discrepant results for the bluemost passband (Papou\v{s}ek $U1$), this figure
shows no evidence for a variation with wavelength of $r_{\rm A}$, $r_{\rm B}$ or $i$. The quantity $u_{\rm A}$ has a weak
dependence on wavelength and in general higher values than expected theoretically.

For the final photometric parameters we have calculated the weighted mean and its reduced $\chi^2$ (\chir)
of the geometrical quantities (i.e.\ those which do not depend on wavelength). As expected given the
discussion above, we find $\chir > 1$ for all quantities. $i$ is least problematic at $\chir = 1.4$
and $k$ is the most difficult at $\chir = 5.9$. This is a very similar situation to that often found
in \reff{studies of transiting planetary systems} \citep{Me08mn,Me10mn,Me12mn}. We have therefore
multiplied the errorbars of the final
weighted-mean values by $\sqrt{\chir}$ to yield reliable errorbars. The large number of available
light curves and careful treatment of uncertainties means that our results are robust despite the
excess \chir\ found for some parameters. Final values of the geometrical parameters are given in
Table\,\ref{tab:lcfit3} and show that $r_{\rm A}$ and $r_{\rm B}$ have been measured to better
than 0.5\% precision.

\begin{figure*} \includegraphics[width=\textwidth,angle=0]{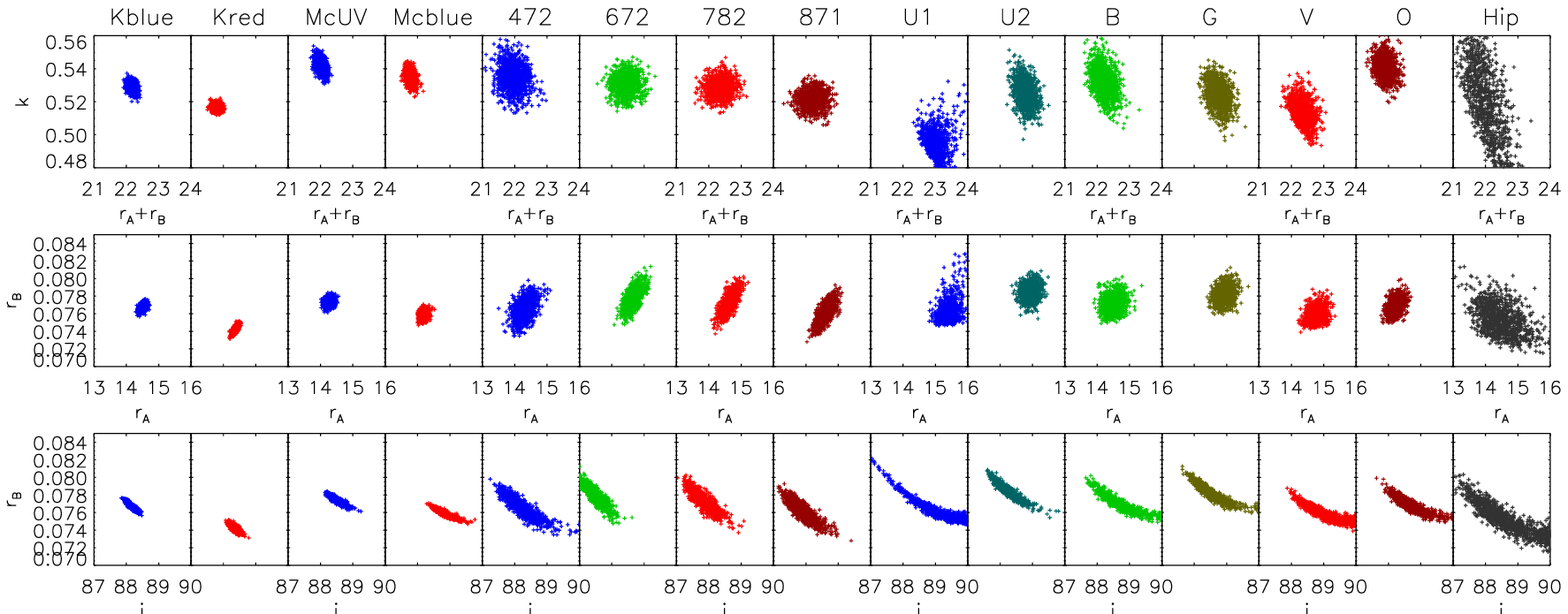}
\caption{\label{fig:mc} Plots of the variations of best-fit parameters
for synthetic datasets generated for the Monte Carlo analysis. From left
to right the panels show the light curves in the order they are given in
Table\,\ref{tab:passbands}. For legibility the x-axis labels for the
parameters $r_{\rm A}+r_{\rm B}$ and $r_{\rm B}$ are only specified for
alternate panels, and the tick values are multiplied by 100 (in effect
omitting the leading ``0.'' in each case). The light curves are colour-coded
according to Figs.\ \ref{fig:plotlc} and \ref{fig:plotlcpap}.} \end{figure*}

Fig.\,\ref{fig:mc} shows the output from Monte Carlo simulations for each light curve and for
the parameters $r_{\rm A}+r_{\rm B}$, $k$, $r_{\rm B}$ and $i$. The points plotted represent
best-fit values for each of the synthetic datasets generated for the Monte Carlo simulations.
It can be seen that the geometric parameters are actually more precisely defined for the oldest
four datasets, and least well-defined for the {\it Hipparcos} survey data. The main parameters
of the fit, $r_{\rm A}+r_{\rm B}$ and $k$, are very weakly correlated (e.g.\ correlation coefficient
$0.04$ for the Utrecht 672 dataset), whereas the derived parameters $r_{\rm A}$ and $r_{\rm B}$
are more correlated ($0.68$). A strong correlation occurs between $i$ and $r_{\rm B}$ ($-0.89$),
as can be seen in the lower panels of Fig.\,\ref{fig:mc}.

\subsection{Continuum light ratio}                                             \label{sec:lc:lr}

The light ratio between the stars is very well determined for YZ\,Cas, \reff{in the passbands for
which we have light curves}, due to it showing total and annular eclipses. For the spectroscopic
analysis in Sect.\,\ref{sec:atmos}, however, we needed a measurement of the continuum
light ratio as a function of wavelength.

We therefore fitted two synthetic spectra from {\sc atlas9} model
atmospheres \citep{Kurucz79apjs} to the light ratios found in the different passbands. For this we ignored
the light curves taken through a wide passband, using only those from \citet{Delandtsheer83aas} and
\citet{Papousek89pbrno}. The synthetic spectra were used only to fill in the gaps between different
passbands, so the choice of their atmospheric parameters was unimportant. Finally, we fitted
a second-order polynomial to the light ratio as a function of wavelength (from 4000 to 7000\,\AA)
to give a smooth and continuous proxy to the true light ratio of the system. Star\,A produces between
86\% and 94\% of the system light in this wavelength region, being more dominant at bluer wavelengths
due to its higher \Teff\ (Fig.\,\ref{fig:lcwave}).


\section{Spectroscopic orbit}                                                   \label{sec:orbit}

\begin{table} \centering \caption{\label{tab:orbit} The orbital elements of YZ\,Cas from
\spd. We fixed the orbital period in our analysis to the value found by \citet{Lacy81apj}.}
\begin{tabular}{l r@{\,$\pm$\,}l r@{\,$\pm$\,}l}
\hline
Parameter                                 & \mc{This work}  & \mc{\citet{Lacy81apj}} \\
\hline
Orbital period $P$ (d)                    & \mc{4.4672235}  & \mc{4.4672235}         \\
Orbital eccentricity $e$                  & \mc{0 (fixed)}  & 0.0 & 0.003            \\
Velocity amplitude $K_{\rm A}$ (\kms)     & 73.05 & 0.19    & 73.35 & 0.35           \\
Velocity amplitude $K_{\rm B}$ (\kms)     & 124.78 & 0.27   & 125.7 & 0.5            \\
Mass ratio $q$                            & 0.585 & 0.002   & 0.583 & 0.008          \\
\hline \end{tabular} \end{table}

The small light contribution of star\,B makes the measurement of its orbital motion and atmospheric
parameters relatively challenging, even from \'echelle spectra. We have therefore used the spectral
disentangling (\spd) approach for our spectroscopic analysis. \spd\ of time-series spectra of binary
star systems \citep{SimonSturm94aa} is a powerful technique which enables the determination of the
optimal set of the orbital parameters of a binary system, and the individual spectra of the components,
simultaneously and self-consistently. A discussion of \spd\ and its practical applications can be found
in reviews by \citet{PavlovskiHensberge10aspc} and \citet{PavlovskiMe12aspc}.

As YZ\,Cas is a dEB, we have followed the analysis approach described in detail by \citet{Hensberge++00aa}.
\reff{\spd\ was performed with the {\sc FDBinary}\footnote{\tt http://sail.zpf.fer.hr/fdbinary} code
\citep{Ilijic+04aspc} which implements the Fourier approach of \citet{Hadrava95aas}.
Since no spectra were obtained during totality in eclipse, there is an ambiguity in the determination of
the continuum level. Therefore, \spd\ was performed in pure separation mode and renormalisation
of disentangled spectra of the individual components to their continua was done with the
light ratio derived from the light curves in Section \ref{sec:lc:lr} \citep{PavlovskiMe12aspc}.}
Since {\sc FDBinary} is based on discrete Fourier transforms there is no limitation on the length
or resolution of the spectra to be analysed, so far as the basic prescriptions for Fourier disentangling
are fulfilled.

There is no indication from photometry (Sections \ref{sec:porb} and \ref{sec:lc}) or previous spectroscopy
\citep{Koch++65apj,Lacy81apj} that the orbit is eccentric. We therefore fixed $e$ to zero in our analysis,
and fitted only for the velocity amplitudes of the two stars ($K_{\rm A}$ and $K_{\rm B}$). Spectral
regions containing Balmer lines were avoided in the optimisation of $K_{\rm A}$ and $K_{\rm B}$, as
\spd\ is very sensitive to minor imperfections in continuum placement over these extremely broad lines.
The large wavelength interval covered by our \'echelle spectra means that there is plenty of spectrum
with a well-defined continuum and containing only metallic lines. \spd\ was performed and  $K_{\rm A}$
and $K_{\rm B}$ calculated for a number of spectral intervals of widths ranging from 50\,\AA\ to 100\,\AA.
The quality of the normalisation and merging of the spectra is reflected in very small standard deviations
of the values of $K_{\rm A}$ ($\pm$0.03\kms) and $K_{\rm B}$ ($\pm$0.04\kms).

A second estimate of the uncertainties in $K_{\rm A}$ and $K_{\rm B}$ can be obtained using other
techniques for least-squares analysis such as bootstrap resampling, but such analysis can require
a lot of computation time. We have therefore performed a jackknife test where a set of solutions
are calculated, each one ignoring a single observed spectrum. We have previously found reasonable
and realistic results using this approach for the dEB AS\,Cam \citep{Pavlovski++11apj}. This was
done for seven of the spectral regions, with the result that the jackknife-derived uncertainties
($\sim$0.2\kms\ and $\sim$0.3\kms) are roughly one order of magnitude greater than the standard
deviation in $K_{\rm A}$ and $K_{\rm B}$ for these spectral regions ($\pm$0.04\kms\ and $\pm$0.05\kms).
We accept the jackknife uncertainties as our final errorbars (Table\,\ref{tab:orbit}); they correspond
to measurement uncertainties of 0.5\% in the masses of the two stars. This is a realistic result for
42 high-quality \'echelle spectra of a dEB containing slowly rotating stars in a circular orbit.


\section{Atmospheric parameters}                                                        \label{sec:atmos}

\begin{figure} \centering \includegraphics[width=\columnwidth]{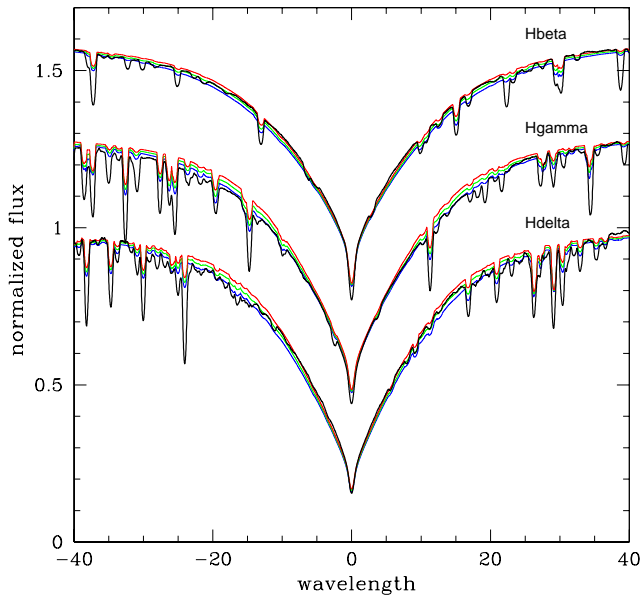} \\
\caption{\label{fig:microprim} Determination of the \Teff\ of \reff{YZ\,Cas\,A}
from its  Balmer line profile. The \reff{renormalised disentangled} profiles are shown
by thin black lines,
and theoretical lines profiles for \Teff s of 9800, 9600 and 9400\,K are shown by red,
blue and green lines. From top to bottom the profiles are of H$\beta$, H$\gamma$ and
H$\delta$. The surface gravity is fixed from the light and RV curve analyses.} \end{figure}

The complexity of spectra of close binary systems due to their composite nature and continuously varying
 Doppler shifts makes the extraction of atmospheric parameters non-trivial. The \spd\ technique is an
important aid to such studies, as it allows the derivation of the individual spectra of the two stars
without needing template spectra for guidance. These separated spectra can then be analysed as if they
were observed spectra of single stars, allowing the determination of the \Teff s and chemical composition
of the two components of the binary system. These numbers are in turn important for the use of dEBs as
distance indicators \citep[e.g.][]{Hensberge++00aa,North++10aa} and as critical tests of stellar
evolutionary theory \citep[e.g.][]{Pavlovski+09mn,Brogaard+11aa}.

The first uses of \spd\ to estimate \Teff s from disentangled individual component spectra were made
for the close binaries DH\,Cep by \citet{SturmSimon94aa} and Y\,Cyg by \citet{Simon++94aa}. These
authors also estimated helium abundances for the components in the systems. On these grounds
\citet{Hensberge++00aa} constructed a self-consistent complementary approach in the analysis
of close binary stars. In a detailed study of the high-mass double-lined dEB V578\,Mon,
\citeauthor{Hensberge++00aa} were able to determine the basic physical properties more accurately
than possible using `standard' techniques. In a follow-up study \citet{PavlovskiHensberge05aa}
presented the first abundance analysis using the broad wavelength range available in disentangled
\'echelle spectra.

In this work we follow the approach of \citet{Hensberge++00aa}. The disentangled spectra have to be
normalised to their intrinsic continua as this cannot be done by \spd. \spd\ essentially attibutes
spectral features to the individual stars according to their orbital motion, and this is not possible
for continuum flux unless spectra were obtained during eclipse. The light factor (LFs) which give the
continuum level of each star can be obtained either from the light curves (using the light ratios from
the best-fitting eclipse model) or by constrained fitting of the Balmer lines of both components
simultaneously \citep{Tamajo++11aa}.

\subsection{Effective temperature and microturbulence}                                  \label{sect:micro}

We proceeded to estimate the \Teff\ of star\,A by fitting its Balmer line profiles and then fine-tuning
the value using the ionisation balance of the many Fe lines in the spectrum. The Balmer lines in A-type
stars are sensitive both to \Teff\ and surface gravity, \logg. In the case of YZ\,Cas, this degeneracy
can be easily sidestepped since our analysis of the light and RV curves results in a measurement of
\logg\ for both stars to an accuracy of better than 0.01\,dex; this is an intrinsic advantage for
spectroscopic analysis of EBs.

Spectral disentangling was performed on three spectral intervals of 150--250\,\AA\ width, centred on
H$\beta$, H$\gamma$ and H$\delta$. In the \spd\ of early-type stars these are the most difficult and
uncertain spectral regions since Doppler shifts due to orbital motion are much smaller than the
intrinsic widths of the Balmer lines.
\reff{In order to minimise systematic errors in the normalisation and merging of \'echelle orders
containing broad Balmer lines we developed a semi-automatic procedure. A high-order polynomial
fit of the blaze function was calculated in adjacent well-defined (`cleaner') \'echelle orders. Then the
blaze function was constructed for \'echelle orders containing broad Balmer lines by interpolation and
scaling between adjacent orders.}
We used the light ratio measured from the photometric analysis
(Sect.\,\ref{sec:lc:lr}) to normalise the disentangled spectra to the intrinsic continuum of each star.
\reff{Since the disentangled (separated) spectra of the components have to be corrected for the
dilution effect, in this particular case by factors of about 1.08 for star\,A, and 14.3
for star\,B, any imperfections in the continuum placement or the determination
of the light ratio would be scaled up by the same amount. Because of the large light ratio
between the components, as well as their atmopsheric properties, in this work the effective
temperature was determined from Balmer lines only for star\,A. It was then used
for deriving the \Teff\ of star\,B through the photometric analysis, and as a
starting point in determination of the \Teff\ of star\,A through detailed
spectroscopic analysis.}

A grid of local thermodynamic equilibrium (LTE) synthetic spectra was calculated using the {\sc uclsyn}
code \citep{Smith92phd,Smalley++01} and {\sc atlas9} model atmospheres for solar metallicity $\MoH = 0$.
The grid covers \Teff s from 8400\,K to 10\,400\,K in steps of 200\,K, for the known surface gravity of
star\,A ($\logg = 3.988$). The \Teff\ of star\,A was determined by minimising the difference between the
reconstructed disentangled spectra and the synthetic spectra in the spectral ranges of H$\delta$
(4070--4170\,\AA), H$\gamma$ (4290--4370\,\AA) and H$\beta$ (4800--4900\,\AA). Minimisation was performed
separately on the blue and red wings. Metallic lines contaminating the Balmer profiles were masked out
during this process. Minimisation from these six wings of Balmer lines gave $\Teff = 9670 \pm 140$\,K
for star\,A, where the errorbar is the 1$\sigma$ error. In Fig.\,\ref{fig:microprim} we show the quality
of fit to the observed (disentangled) spectra of the Balmer lines by the synthetic spectra.

In the spectra of the components several species appear in two or even three ionization stages, e.g.\ Si
for star\,A. The most numerous lines in the spectra of both components are \ion{Fe}{i} and \ion{Fe}{ii}
lines, and they are well suited to the determination of both \Teff\ and microturbulent velocity (\micro).
\Teff\ was determined from \ion{Fe}{i} and \ion{Fe}{ii} lines under the condition that abundances should
not depend on ionisation state or excitation potential \citep{Gray08book}. We determined \micro\ by
requiring the Fe abundance to be independent of equivalent width (EW) \citep{Magain84aa}.

The measurements of EWs \reff{were obtained by direct integration of the line profiles and
abundances for \ion{Fe}{i} and \ion{Fe}{ii} calculated using {\sc uclsyn}.}  We have
followed the critical selection of \ion{Fe}{i} and \ion{Fe}{ii} lines from \citet{Qiu+01apj} in deriving
\Teff\ and \micro, but all identified \ion{Fe}{} lines were measured. The lines were selected on the
basis of the reliability of their atomic data, mostly {\em gf} values, and their strength. We agree
with \citet{Qiu+01apj} that including all lines, most of which are weak lines with EW $< 5$ m\AA,
results in a larger scatter and less reliable abundances. In total we have 51 \ion{Fe}{i} and 30
\ion{Fe}{ii} reliable lines for star\,A (see Table\,\ref{tab:abundfin}), with EWs ranging from 5
to 100\,m\AA. For the cooler star\,B there are many more Fe lines, 216 \ion{Fe}{i} and 24 \ion{Fe}{ii},
and these have EWs of 5--70\,m\AA. All Fe lines detected were checked against new entries for {\em gf}
values in the {\sc vald}\footnote{http://ams.astro.univie.ac.at/vald} line lists \citep[][and references
therein]{Kupka+00balta}.

For star\,A we found $\Teff = 9520 \pm 120$\,K, in acceptable agreement with the value from fitting
the Balmer lines, and $\micro = 3.8 \pm 0.1$\kms. In the case of star\,B, we estimated a preliminary
\Teff\ from its colour indices because its Balmer lines are much weaker. We refined the value using
the ionisation balance of Fe, finding $\Teff = 6880 \pm 240$\,K and $\micro = 1.7 \pm 0.1$\kms. The
observed spectra are quite sensitive to \micro\ so we have been able to constrain the values well.
The derived value for star\,A is typical \reff{for Am stars \citep[e.g.][]{Landstreet+09aa} whilst
the value for star\,B fits the trend found for stars of similar \Teff\ \citep[e.g.][]{Bruntt+10mn,Doyle+13mn}.}

\citet{DelandtsheerMulder83aa} estimated the \Teff\ of star\,A from an IUE low-resolution calibrated
spectrum, in a determination also involving the distance and line-of-sight interstellar extinction for
YZ\,Cas. The authors concluded that $\Teff = 10\,600 \pm 400$\,K, arguing against the value of $9000$\,K
resulting from visual spectral classification \citep{Hill+75mmras}. \citet{Ribas+00mn} used
intermediate-band photometry and grids of spectra from model atmospheres to revise the \Teff\
scale using a sample of dEBs. They found $\Teff = 9100 \pm 300$\,K for star\,A and $\Teff = 6600 \pm 250$\,K
for star\,B. Our spectroscopic results, based both on Balmer line profiles and Fe ionisation balance,
corroborate their estimates to within 1$\sigma$.

\subsection{Abundances}

\begin{table} \centering \caption{\label{tab:abundfin} Measured chemical abundances
in the photospheres of the component stars of YZ\,Cas, derived in the LTE approximation.
$N$ is the number of lines used for each element.}
\setlength{\tabcolsep}{3pt}
\begin{tabular}{l r r@{\,$\pm$\,}l r@{\,$\pm$\,}l r r@{\,$\pm$\,}l r@{\,$\pm$\,}l}
\hline
Ion & \multicolumn{5}{c}{YZ Cas A} & \multicolumn{5}{c}{YZ Cas B} \\
\   & $N$ & \mc{$\log\epsilon{\rm (X)}$} & \mc{[X/H]}
    & $N$ & \mc{$\log\epsilon{\rm (X)}$} & \mc{[X/H]} \\
\hline
\ion{C}{}   & 55 & \reff{8.85} & 0.10 &    0.36 & \reff{0.11}  &  29 &  8.49 & 0.06 &  0.06   & \reff{0.08} \\
\ion{O}{}   & 25 & \reff{9.03} & 0.08 &    0.34 & \reff{0.09}  &  10 &  8.89 & 0.13 &  0.20   & \reff{0.14} \\
\ion{Na}{}  &  5 & \reff{6.99} & 0.08 &    0.68 & \reff{0.09}  &   8 &  6.34 & 0.08 &  0.10   & \reff{0.10} \\
\ion{Mg}{}  & 19 & \reff{7.91} & 0.09 &    0.26 & \reff{0.10}  &  14 &  7.52 & 0.14 & $-$0.08 & \reff{0.15} \\
\ion{Al}{}  &  2 & \reff{7.26} & 0.09 &    0.73 & \reff{0.09}  &     &     \mc{ }   &       \mc{ }          \\
\ion{Si}{}  & 63 & \reff{7.89} & 0.08 &    0.31 & \reff{0.09}  &  46 &  7.48 & 0.05 & $-$0.03 & \reff{0.07} \\
\ion{S}{}   & 22 & \reff{7.83} & 0.12 &    0.65 & \reff{0.12}  &  13 &  7.26 & 0.04 &   0.14  & \reff{0.05} \\
\ion{Ca}{}  & 29 & \reff{6.66} & 0.11 &    0.27 & \reff{0.12}  &  25 &  6.37 & 0.11 &   0.03  & \reff{0.12} \\
\ion{Sc}{}  & 13 & \reff{3.01} & 0.11 & $-$0.19 & \reff{0.12}  &  10 &  3.11 & 0.07 & $-$0.04 & \reff{0.08} \\
\ion{Ti}{}  & 70 & \reff{5.37} & 0.11 &    0.34 & \reff{0.12}  & 120 &  4.97 & 0.12 &   0.02  & \reff{0.13} \\
\ion{V}{}   &  9 & \reff{4.45} & 0.09 &    0.47 & \reff{0.12}  &  13 &  4.10 & 0.13 &   0.17  & \reff{0.15} \\
\ion{Cr}{}  & 71 & \reff{6.24} & 0.09 &    0.53 & \reff{0.10}  & 127 &  5.67 & 0.11 &   0.03  & \reff{0.12} \\
\ion{Mn}{}  & 22 & \reff{5.91} & 0.12 &    0.39 & \reff{0.11}  &  23 &  5.40 & 0.10 & $-$0.03 & \reff{0.11} \\
\ion{Fe}{}  & 81 & \reff{8.12} & 0.10 &    0.54 & \reff{0.11}  & 240 &  7.51 & 0.10 &   0.01  & \reff{0.11} \\
\ion{Co}{}  &    &          \mc{ }    &      \mc{ }            &  44 &  5.01 & 0.13 &   0.02  & \reff{0.15} \\
\ion{Ni}{}  &  8 & \reff{6.96} & 0.09 &    0.67 & \reff{0.10}  & 109 &  6.25 & 0.11 &   0.03  & \reff{0.12} \\
\ion{Cu}{}  &  1 & \reff{4.70} & 0.13 &    0.43 & \reff{0.14}  &   9 &  4.19 & 0.11 &   0.00  & \reff{0.12} \\
\ion{Zn}{}  &  1 & \reff{5.80} & 0.13 &    1.17 & \reff{0.14}  &   2 &  4.60 & 0.17 &   0.04  & \reff{0.18} \\
\ion{Y}{}   &  6 & \reff{3.25} & 0.11 &    0.96 & \reff{0.12}  &  14 &  2.26 & 0.08 &   0.05  & \reff{0.09} \\
\ion{Zr}{}  &  5 & \reff{3.64} & 0.14 &    0.99 & \reff{0.16}  &   8 &  2.58 & 0.12 &   0.00  & \reff{0.13} \\
\ion{Ba}{}  &  2 & \reff{3.43} & 0.15 &    1.15 & \reff{0.18}  &   2 &  2.19 & 0.12 &   0.01  & \reff{0.15} \\
\ion{La}{}  &    &          \mc{ }    &      \mc{ }            &  19 &  1.29 & 0.10 &   0.19  & \reff{0.11} \\
\ion{Ce}{}  &    &          \mc{ }    &      \mc{ }            &  26 &  1.58 & 0.11 &   0.00  & \reff{0.12} \\
\ion{Pr}{}  &    &          \mc{ }    &      \mc{ }            &   4 &  0.72 & 0.12 &   0.00  & \reff{0.13} \\
\ion{Nd}{}  &    &          \mc{ }    &      \mc{ }            &  31 &  1.56 & 0.10 &   0.14  & \reff{0.11} \\
\ion{Sm}{}  &    &          \mc{ }    &      \mc{ }            &   2 &  1.08 & 0.17 &   0.12  & \reff{0.17} \\
\hline \end{tabular} \end{table}

\begin{figure} \centering
\includegraphics[width=\columnwidth]{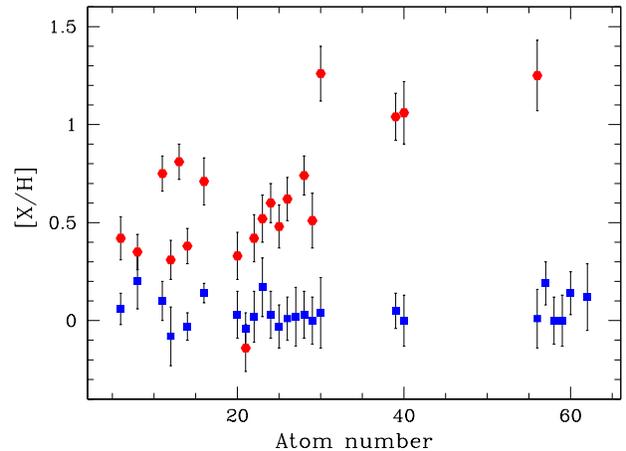}
\caption{\label{fig:abundfin} Comparison between abundances derived for
star\,A (red filled circles) and star\,B (blue filled squares).} \end{figure}

For stars with \Teff s similar to those of the components of YZ\,Cas, a large number of spectral lines
arise from neutral and singly ionised species. The number of good lines for abundance determination
is, however, decreased due to the rotation of the component stars. We selected good lines based on the
quality of their atomic data and the amount of blending with other lines.

We used {\sc uclsyn} to measure EWs and calculate the abundances. Table\,\ref{tab:abundfin} contains
the derived abundances for all elements identified in the spectrum of either star, in two forms. Firstly,
the number density $\epsilon$ is given on a scale where $\log\epsilon{\rm (H)} = 12.0$. Secondly,
the value for element X is converted into logarithmic abundance versus the Sun ([X/H]) using the
standard solar abundances from \citet{Asplund+09araa}, the most recent critical evaluation of the
solar abundances. The errorbars on these quantities are the r.m.s.\ errors of the results for individual
 lines, so are indicative of the number and quality of the lines used rather than the true uncertainties.
Additional contributions to the uncertainties arise from the quality of the atomic data, the quality of
the spectra and their normalisation, and uncertainties in the measured \Teff, \logg\ and \micro\ values.
The spectra are of high quality and their normalisation is well-defined away from the Balmer lines, so
these should not give rise to significant systematic errors. Similarly, the atmospheric parameters are
measured to high precision so will not cause much uncertainty.

The abundances in Table\,\ref{tab:abundfin} are visualised in Fig.\,\ref{fig:abundfin}. We find that the
chemical abundances for star\,B are close to solar; \reff{a calculation of the metallicity from the abundances
listed in Table\,\ref{tab:abundfin} for star\,B and replacing missing elements with abundances
given by \citet{Asplund+09araa} returns a value of $Z = 0.0170 \pm 0.024$ (where $Z$ is the mass fraction of metals)
which is slightly more than 1$\sigma$
away from the solar value of $Z = 0.0134$ derived in \citet{Asplund+09araa}.} These
values should be appropriate for the bulk
chemical composition of the system as a whole because the photosphere is expected to represent the
internal composition for a 6880\,K dwarf star. The abundances for star\,A are all significantly
super-solar except for Sc, confirming the Am nature of this star. Zn, Y, Zr and Ba are all
overabundant by 1\,dex relative to solar. We have not been able to measure the abundances of any
rare-earth elements, even though a strong over-abundance of these is a hallmark of the Am phenomenon
\citep{Wolff83book}.

\reff{We calculated a provisional set of abundances for star\,A using model atmospheres of solar abundance rather than abundances tuned to the true metallicity of the star. For our final abundance measurements (Table\,\ref{tab:abundfin}), we used model atmospheres with scaled-solar metallicity and a metal abundance of [X/H] $=$ 0.5. The use of scaled-solar model atmospheres should be sufficient for our work because the blanketing effect on the atmospheric structure is mostly due to iron-group elements. \citet{Smalley93mn} has shown the effect on the derived abundances to be small: a few hundreths of a dex in abundance and probably no more than 100\,K in \Teff. We found that the use of model atmospheres with [X/H] $=$ 0.5 rather than solar abundance had the effect of increasing the derived abundances for all elements by between 0.05 and 0.10\,dex, with the exception of oxygen for which the measured abundance increased by 0.003\,dex, only.}

There exists one previous abundance study of YZ\,Cas, by \citet{DelandtsheerMulder83aa}. These authors
derived the atmospheric properties and the abundances of six elements from {\reff{IUE} high-resolution spectra.
They adopted a higher \Teff\ (10\,600\,K) and measured a higher \micro\ ($5.8 \pm 1.7$\kms) than we find,
and this is probably why they derived exceptionally high abundance values (from 0.80\,dex for Fe to
3.4\,dex for Co relative to solar). We do not confirm these overabundances for five of the six chemicals
(Si, Cr, Mn, Fe, Ni). We did not obtain an abundance for the sixth, Co, but the value from
\citet{DelandtsheerMulder83aa} is almost certainly also wrong.

\subsection{Rotational velocity}                                                    \label{sect:vrot}

The disentangled spectra of YZ\,Cas have S/N values of about 1500 and 200 for star\,A and star\,B,
respectively, and are rich in information. We first measured the instrumental broadening profile using
both ThAr emission lines and and telluric absorption lines, finding FWHM $=$ 0.13\,\AA. Despite both stars
having only moderate rotational velocities, the lines in their spectra are strongly overlapping. We
therefore measured $v\sin i$ by line-profile fitting of complex blends using the {\sc uclsyn} code.

In each disentangled spectrum several spectral regions were selected containing lines with good atomic
data. Line-profile fitting then yielded $v_{\rm A} \sin i = 29.2$\kms\ and $v_{\rm B} \sin i = 15.0$\kms.
The formal errors on these determinations are $\pm$0.1\kms\ and $\pm$0.3\kms, respectively. Additional
contributions to these uncertainties come from the instrumental broadening, its variation with wavelength,
continuum normalisation, microturbulence and macroturbulence. We therefore adopt larger errorbars of
$\pm$0.5\kms\ for both \vsini\ measurements.

If both components show synchronous rotation the ratio of their velocities should equal the ratio of
their radii. We find $v_{\rm A}\sin i / v_{\rm B}\sin i = 0.514 \pm 0.019$ and $k = 0.5246 \pm 0.0026$,
so these values are consistent to within 0.5$\sigma$. This matches our expectation that the components
should be rotating synchronously with the orbital motion.


\section{The physical properties and distance of YZ Cassiopeiae}             \label{sec:absdim}

\begin{table} \label{tab:absdim}
\caption{The physical properties of the YZ\,Cas system.}
\begin{center} \begin{tabular}{l r@{\,$\pm$\,}l r@{\,$\pm$\,}l} \hline
Parameter                     &    \mc{Star A}   &    \mc{Star B}    \\ \hline
Orbital separation (\Rsun)    &\multicolumn{4}{c}{$17.468 \pm 0.029$}\\
Mass (\Msun)                  &  2.263  & 0.012  &  1.325  & 0.007   \\
Radius (\Rsun)                &  2.525  & 0.011  &  1.331  & 0.006   \\
$\log g$ [cm\,s$^{-2}$]       &  3.988  & 0.004  &  4.311  & 0.004   \\
\Vsync\ (\kms)                &  28.61  & 0.12   &  15.08  & 0.07    \\
\Vsini\ (\kms)                &  29.2   & 0.5    &  15.0   & 0.5     \\[3pt]
\Teff (K)                     &  9520   & 120    &  6880   & 240     \\
$\log(L/\Lsun)$ $^1$          &  1.672  & 0.022  &  0.552  & 0.061   \\
\Mbol\ $^1$                   &  0.57   & 0.06   &  3.37   & 0.15    \\
Distance (pc)                 &  \multicolumn{4}{c}{$103.8 \pm 1.3$} \\
\hline \end{tabular} \end{center}
$^1$ Calculated assuming $\Lsun = 3.844${$\times$}10$^{26}$\,W
\citep{Bahcall++95rvmp} and $\Mbol\sun = 4.75$ \citep{Zombeck90book}.
\end{table}

Using the photometric and spectroscopic results from Sections \ref{sec:lc} and \,\ref{sec:orbit}, we have
calculated the physical properties of the YZ\,Cas system (Table\,\ref{tab:absdim}). This was done using
the {\sc jktabsdim} code \citep{Me++05aa}, which propagates uncertainties via a perturbation analysis.
\reff{We adopted the physical constants tabulated by \citet{Me11mn}.} The masses and radii are measured to precisions of
0.5\%, representing a useful improvement over previous analyses. The high precision of these determinations
is due to several circumstances. Firstly, we had access to extensive observational data: 15 published light
curves plus 42 new high-quality \'echelle spectra. Secondly, the intrinsic character of the system makes it
well-suited to detailed analysis. The total eclipses mean the light curve solutions are very well-defined,
and the slow rotation and large number of spectral lines means the orbital velocities of the two components
are measurable to high precision. Compared to \citet{Lacy81apj}, we find very similar radii but masses
smaller by 2--3$\sigma$.

We have also measured the atmospheric parameters of both stars to high precision. Our new \Teff\ measurements
are precise and support previous optical rather than UV determinations. The rotational velocities of both
stars are consistent with synchronocity, as expected from the timescales for rotational synchronisation
\citep[see][]{Zahn75aa,Zahn77aa}.

From the known masses and radii and the \Teff s obtained in Sect.\,\ref{sec:atmos} we have calculated the
luminosities (expressed logarithmically with respect to the Sun in Table\,\ref{tab:absdim}) and absolute
bolometric magnitudes of the two stars. From these we have measured the distance to the system using
empirical surface brightness relations (see \citealt{Me++05aa} and \citealt{Kervella+04aa}) and by the
usual method involving bolometric corrections \citep[e.g.][]{Me++05iauc}. For this process we adopted
apparent magnitudes in the $B$ and $V$ bands from \citet{Hog+97aa} and in the $JHK_s$ bands from
\citet{Skrutskie+06aj}. Tabulated bolometric corrections were taken from \citet{Bessell++98aa} and
\citet{Girardi+02aa}. All distance measurements agree well, and we take the $K_s$-band
surface-brightness-based distance of $103.8 \pm 1.3$\,pc as our final value.

The good agreement between the optical ($BV$) and IR ($JHK_s$) distance measurements argues against
the presence of significant interstellar absorption; by requiring consonant distances we find an upper
limit of only $E(B-V) = 0.03$\,mag (3$\sigma$). We have also interpreted the light ratio derived from
the \citet{Papousek89pbrno} $V$ light curve as a light ratio in the Johnson $V$ band to obtain a separate
 distance estimate for the two stars. These agree to within their errorbars, so the \Teff s and radii
in Table\,\ref{tab:absdim} pass this consistency check.

Both stars in YZ\,Cas are situated outside the \reff{$\delta$ Sct} instability strip
\citep{Dupret+05aa,Uytterhoeven+11aa}:
star\,A is hotter than the blue edge of the fundamental radial mode, and star\,B is cooler than its red
edge. With a \Teff\ of 9520\,K, star\,A is among the hottest of the Am stars.


\section{Comparison with theoretical models}                                           \label{sec:models}

\begin{figure} \centering \includegraphics[width=\columnwidth]{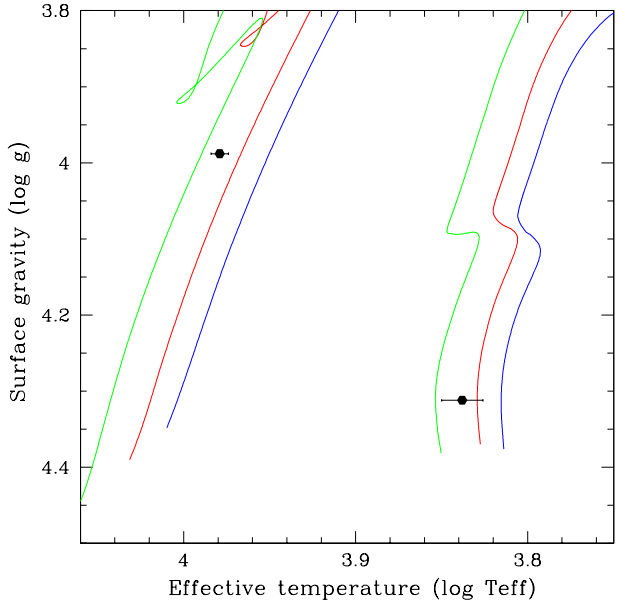} \\
\caption{\label{fig:tracks} The locations of the components of YZ\,Cas in the
$\log\Teff$ versus \logg\ plane. The Geneva evolutionary tracks \citep{Mowlavi+12aa}
for their masses are plotted for the metallicity values $Z = 0.006$ (green lines),
0.010 (red lines) and 0.014 (blue lines).} \end{figure}

Considerable attention has been devoted to the metallicity of the component stars of YZ\,Cas. The relatively
large difference in their masses means they could be a stringent test of the stellar evolutionary models
\citep{LastennetVallsgabaud02aa}. \reff{These authors found a problem in the} discrepancy in metallicity estimated
from matching the physical properties of the system to the predictions of theoretical stellar models
($Z = 0.015$), and the metallicity derived for star\,A from
UV spectroscopy ($Z = 0.036 \pm 0.005$; \citealt{DelandtsheerMulder83aa}). \citet{Lastennet++01assl}
challenged the UV metallicity determination on the basis that photometric methods do not indicate
a high $Z$ and are in fact compatible with solar metallicity.

Our detailed abundance analysis based on optical \'echelle spectra has revealed the photospheric chemical
 composition of both stars. Star\,A shows a high metallicity which cannot be taken as an indication of its
 internal composition. Star\,B, however, shows a solar metallicity which should be representative of its
bulk metal abundance.

We have compared the masses, radii and \Teff s of the components of YZ\,Cas to tabulated predictions from
several sets of theoretical stellar evolutionary models. An immediate results is that the best fit to the
observed properties is found for a {\it subsolar} metal abundance, whereas a solar metallicity results in
predicted \Teff s which are too small to match the observed values. We also were able to infer a precise age,
 $\tau$, for the system. The Teramo models \citep{Pietrinferni+04apj} give a good fit for $Z = 0.010$ and
$\tau = 490$\,Myr, with a slight preference for models with core overshooting over canonical models.
The VRSS models \citep{Vandenberg++06apjs} agree very well with the measured properties for $Z = 0.010$
and $\tau = 545$\,Myr, with an acceptable fit also being found for $Z = 0.0125$ and $\tau = 530$\,Myr.
The PARSEC models \citep{Bressan+12mn} for $Z = 0.010$ and $\tau = 545$\,Myr are almost identical to the
VRSS ones, so also fit well. Finally, an investigation of the Y$^2$ models \citep{Demarque+04apjs} resulted
in a good fit for $Z = 0.01$ and $\tau = 550$\,Myr, where a higher $Z$ causes the predicted \Teff s to become
too low and a lower $Z$ underpredicts the radius of star\,B.

In Fig.\,\ref{fig:tracks} we show an alternative approach where evolutionary tracks interpolated to the
specific masses of the two stars are plotted in the $\log\Teff$ versus \logg\ diagram for three metallicities.
The models used are the most recent versions from the Geneva group \citep{Mowlavi+12aa}, which have modest
convective core overshooting and do not account for stellar rotation. \reff{We found the best-fitting metallicities and ages
to be $Z = 0.008 \pm 0.001$ and 420\,Myr for star\,Am and $Z = 0.009 \pm 0.002$ and 670\,Myr for star\,B.} This corroborates our
inferences from the four sets of models discussed above.

We therefore find that the physical properties of both stars match the predictions of theoretical models
for a metallicity of $Z = 0.009 \pm 0.003$ and an age of 400--700\,Myr. This is troubling because the
photospheric abundances we find for star\,B show it to have an approximately solar chemical composition.
We coclude that either the theoretical models are incorrect or the photospheric abundances of star\,B
do not represent its bulk chemical composition.


\section{Conclusions}

We have presented a detailed analysis of the dEB YZ\,Cas based on 15 published light curves and 42 new
high-quality \'echelle spectra. The principal attraction of this object is that it contains two quite
different stars: one an A-type metallic-lined star and the other a smaller and less massive F2 dwarf.
We were therefore able to investigate the Am nature of star\,A using the chemically normal star\,B as
a reference.

We have measured the masses and radii of both component to a precision of 0.5\%, due to the large amount
of observational material as well as the co-operative nature of the system. The time-resolved spectra
were analysed using spectral disentangling, allowing us to derive the individual spectra of the two
stars as well as their velocity amplitudes. From the separated spectra we obtained the atmospheric
parameters (\Teff\ and \micro) and photospheric chemical compositions of both stars.

Star\,A shows clear signs of the Am phenomenon: enhanced metals, depleted Sc, and over-abundances of
Zn, Y, Zr and Ba by 1\,dex. By contrast, star\,B shows normal chemical abundances which are consistent
with a solar or slightly super-solar chemical composition. Whilst the Am phenomenon is a `surface disease',
the abundances of star\,B should represent the bulk chemical composition of both stars. It is therefore
surprising that theoretical stellar evolutionary models require a significantly sub-solar metallicity to
reproduce the properties of the YZ\,Cas system. Our results cannot differentiate between the possibilities
that the model predictions are wrong and that the photospheric abundances of star\,B do not represent
the true chemical composition of either star.

The Am stars are defined phenomenologically: in classification-resolution spectra the Ca K line is
appropriate for an early-A spectral type and metallic lines point to a late-A to F type, whilst the
hydrogen Balmer lines are intermediate \citep{Roman++48apj}. With the widespread use of higher-resolution
spectra a new definition of the Am phenomenon has evolved. \citet{Conti70pasp} defined it as an apparent
surface underabundance of Ca and/or Sc, and/or an apparent overabundance of Fe-group and heavier elements.
Modern spectroscopic studies have revealed much variety in observed abundance patterns, and difficulties
exist in delineating the border between normal and chemically peculiar A stars. Interesting discussions
of the Am phenomenon in can be found in \citet{AdelmanUnsuree07balta} and \citet{Murphy+12mn}. While
Am stars were long considered to have quiet atmospheres that were therefore not pulsating, high-precision
 photometry from both ground-based (SuperWASP) and space-based ({\it Kepler}) surveys have yielded the
detection of more than 200 A and F dwarfs with Am characteristics and  detectable pulsations
\citep{Smalley+11aa2,Balona+11mn}. Important constraints of the origin of the Am phenomenon are also
provided by their rotation and binarity \citep{Fossati+08aa,Stateva++12mn}.

Our work on YZ\,Cas has yielded precise measurements of the mass, radius and \Teff\ of an Am star,
plus the abundances of 26 elements in its photosphere. Its F2\,V companion has a solar chemical
composition which should reflect the internal composition of both stars. Theoretical models cannot
reproduce the physical properties of the stars for this composition. Further work is needed to
understand the nature of this discrepancy and to model the processes occurrent in the atmospheres
of Am stars.


\section*{Acknowledgments}

We are grateful to Frank Verbunt, Miloslav Zejda, Robert Van Gent and Jerzy Kreiner for discussions and
information. We thank the referee, Colin Folsom, for a helpful report.
JS acknowledges financial support from STFC in the form of an Advanced Fellowship. Based
on observations made with the Nordic Optical Telescope, operated on the island of La Palma jointly by
Denmark, Finland, Iceland, Norway, and Sweden, in the Spanish Observatorio del Roque de los Muchachos
of the Instituto de Astrof\'{\i}sica de Canarias. The following internet-based resources were used in
research for this paper: the ESO Digitized Sky Survey; the NASA Astrophysics Data System; the SIMBAD
database operated at CDS, Strasbourg, France; and the ar$\chi$iv scientific paper preprint service
operated by Cornell University.


\bibliographystyle{mn_new}

\end{document}